\documentclass[preprint,12pt]{elsarticle}

\usepackage{amssymb}

\biboptions{comma,compress}
\usepackage{amsmath}
\usepackage{graphicx}
\usepackage{dcolumn}
\usepackage{bm}
\usepackage{hyperref}

\usepackage{color}
\usepackage{blindtext}
\usepackage[printonlyused]{acronym}
\usepackage{url}
\usepackage[nowatermark]{fixmetodonotes}
\usepackage[margin=1.in]{geometry}
\usepackage{floatrow}
\usepackage{listings}

\newcounter{bla}

\journal{Computer Physics Communications}

\begin{document}

\begin{frontmatter}



\title{A universal implementation of radiative effects in neutrino event generators}


\author[TAU]{J\'{u}lia Tena-Vidal}
\author[TAU]{Adi Ashkenazi}
\author[ODU]{L.~B.~Weinstein}
\author[UoM]{Peter Blunden}
\author[Pitt]{Steven Dytman}
\author[Fermilab]{Noah Steinberg}

\address[TAU]{Tel Aviv University, Tel Aviv 69978, Israel}
\address[ODU]{Old Dominion University, Norfolk, VA, USA}
\address[UoM]{University of Manitoba, Winnipeg, MB R3T 2N2, Canada}
\address[Pitt]{University of Pittsburgh, Department of Physics and Astronomy, Pittsburgh, PA 15260, USA}
\address[Fermilab]{Fermi National Accelerator Laboratory, Batavia, IL, USA}


\begin{abstract}
Due to the similarities between electron-nucleus ($eA$) and neutrino-nucleus scattering ($\nu A$), $eA$  data can contribute key information to improve cross-section modeling in $eA$ and hence in $\nu A$  event generators.
However, to compare data and generated events, either the data must be radiatively corrected or radiative effects need to be included in the event generators.
We implemented a universal radiative corrections program that can be used with all reaction mechanisms and any $eA$ event generator.  
Our program includes real photon radiation by the incident and
scattered electrons, and virtual photon exchange and photon vacuum polarization diagrams.
It uses the ``extended peaking" approximation for electron radiation and neglects charged hadron radiation. This method, validated with GENIE, can also be extended to simulate $\nu A$ radiative effects. This work facilitates data-event-generator comparisons used to improve $\nu A$ event generators for the next-generation of neutrino experiments. 
\end{abstract}

\begin{keyword}
Electron-Scattering; Radiative Effects; Neutrino-Scattering; 

\end{keyword}
\end{frontmatter}
\section{Introduction}

Future long-baseline neutrino oscillation experiments, such as the Deep Underground Neutrino Experiment (DUNE)~\cite{Falcone:2022upq}, aim to measure neutrino oscillation parameters with unprecedented precision. 
Such precision demands correspondingly precise reconstruction of the incoming neutrino flux as a function of neutrino energy, which relies on comprehensive neutrino-nucleus ($\nu A$) cross-section models, typically encapsulated in event generators. In the absence of a comprehensive theoretical description of all $\nu A$ scattering, these event generators rely  on a combination of  nuclear theory, $\nu A$ data~\cite{PhysRevD.104.072009},  and electron-nucleus ($eA$) data~\cite{PhysRevD.103.113003,PhysRevC.100.054606}.

Extracting cross-section information from the $\nu A$ data is complicated by the broad energy spread of neutrino beams and by low statistics.  
However, due to the similarities between $\nu A$ and $eA$ interactions, we can use $eA$ data to  constrain $\nu A$ cross section models.  
Neutrinos interact via weak vector and axial-vector currents, whereas for electrons the electromagnetic vector current interaction dominates.
Both can interact through quasielastic scattering (QEL) on single nucleons, two-body currents knocking out pairs of nucleons (MEC), exciting single nucleons to a $\Delta(1232)$ or higher resonance (RES) or by deep inelastic scattering from a quark in the nucleon (DIS).
Both $eA$ and $\nu A$ require precise knowledge of the nuclear ground state and outgoing hadrons undergo final-state interactions (FSI) in the residual nucleus. Several collaborations are working to provide $eA$ data~\cite{PhysRevD.103.113003,Khachatryan:2799244,JeffersonLabHallA:2020rcp,JeffersonLabHallA:2022cit,JeffersonLabHallA:2022ljj,Ankowski_2023}.

\begin{figure}[ht]
    \includegraphics[width=\columnwidth]{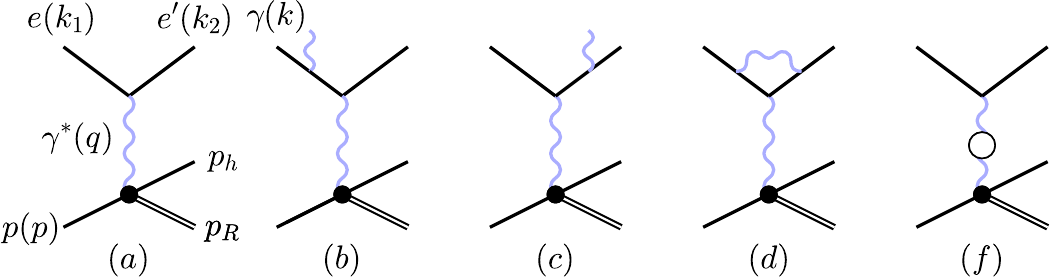}
    \caption{\label{fig:diagrams} Target-independent gauge-invariant set of radiative correction diagrams.  (a) The incident electron $e(k_1)$ exchanges a virtual photon $\gamma^*(q)$ with the nucleus $p(p)$, with a scattered electron $e'(k_2)$ and a nuclear final state (shown as $p_h$ and $p_R$).  (b) and (c)  emission of a real photon by the incident and scattered electrons, respectively.  (d) exchange of a virtual photon between the incident and scattered electrons (vertex correction and (f) virtual  pair production (vacuum polarization).  Not shown is the exchange of a low energy second virtual photon from one of the electrons to the nucleus or the radiation of a real photon by the nucleus or one of the hadrons. }
\end{figure}

Electron-nucleus scattering is significantly affected by radiative effects and QED nuclear medium effects~\cite{Tomalak:2022kjd,Tomalak:2023kwl,Tomalak:2024lme}.
The incoming and scattered electrons, as well as any emitted charged hadron, can radiate  real and virtual photons, see Fig.~\ref{fig:diagrams}.  Real photon radiation, Fig.~\ref{fig:diagrams}(b-c), changes the observed event kinematics and hence the cross section by changing  the incident and/or scattered electron energy and angle. Virtual photon radiation, Fig.~\ref{fig:diagrams}(d-f) changes the  cross section but not the event kinematics~\cite{PhysRev.76.790,RevModPhys.46.815,Vanderhaeghen_2000}.  In charged-current (CC) neutrino scattering, the incident neutrino does not radiate. However, the final-state charged lepton exchanges virtual and real photons, leading to radiative effects of a similar magnitude to those observed in electron scattering~\cite{Tomalak:2021hec,Tomalak:2022xup}.

Neutrino event generators aim to simulate $\nu A$ interactions for all interaction mechanisms and targets. 
Some neutrino event generators, such as \texttt{GENIE}~\cite{andreopoulos2015genie,PhysRevD.103.113003} and \texttt{GIBUU}~\cite{GiBUU}, can also simulate $eA$ interactions with the analogous interaction models.
They aim to utilize consistent nuclear, interaction, and FSI models for both electrons and neutrinos.
Other neutrino event generators such as \texttt{NEUT}~\cite{Hayato_2021,NEUT}, \texttt{NuWRO}~\cite{GOLAN2012499} and \texttt{Achilles}~\cite{Isaacson_2023} can describe $\nu A$ and $eA$ QEL interactions and are working towards the implementation of $eA$ RES and DIS models. None of the existing neutrino event generators account for either $\nu A$ or $eA$ radiative effects.

Published $eA$ cross sections are typically corrected for radiative effects using event generators. The procedure to compute radiative effects for inclusive electron scattering was first derived by J.~Schwinger~\cite{PhysRev.76.790} and was later modified by Mo and Tsai for $(e,e'p)$ reactions~\cite{RevModPhys.46.815}.
A compilation of radiative correction programs can be found at the Jefferson Laboratory (JLAB) website~\cite{JLABRad}. An widely used code is the \texttt{SIMC} Monte Carlo, used in Hall C/A analyses~\cite{SIMC}. However, existing radiative correction codes, such as \texttt{SIMC}, are usually process, topology specific and are restricted to a limited phase space which matches the spectrometer acceptance. Hence, they cannot be used to correct radiative effects for more general semi-inclusive and exclusive $eA$ measurements, which include a wide range of reaction mechanisms. 

Incorporating radiative effects in event generators is difficult and generator dependent.
Radiative effects are usually accounted for by applying weights to generated weighted events, but this is not supported by all available event generators.  
The Electrons for Neutrinos collaboration ($e4\nu$) is analyzing $eA$ data~\cite{PhysRevD.103.113003,Khachatryan:2799244} using the large acceptance  CLAS6 and CLAS12~\cite{CLAS:2003umf,BURKERT2020163419} spectrometers at the Thomas Jefferson National Accelerator Facility.
While there is a lot of inclusive $eA$ data, there is very little hadron electroproduction data~\cite{JeffersonLabHallA:2022ljj,JeffersonLabHallA:2020rcp,JeffersonLabHallA:2022cit}.
 $e4\nu$ focused on quasi-elastic like signatures~\cite{Khachatryan:2799244,Dytman24}. 
Ref.~\cite{Khachatryan:2799244} used Ref.~\cite{PhysRevC.64.054610} approach to account for radiative effects in QEL events using GENIE.
Looking ahead, $e4\nu$ is  working towards a comprehensive electron scattering cross-section library that includes more complex final states and nuclear targets with direct impact on the neutrino community.
In order to extend Ref.~\cite{Khachatryan:2799244}'s implementation to all interaction mechanisms and experimental conditions, radiative effects must be implemented in a more general way.

In this paper we propose a general approach based on Ref.~\cite{PhysRevC.64.054610,Vanderhaeghen_2000} to implement radiative effects for $eA$ event generators without modifying the generators themselves. 
It accounts for radiation of a real photon by the electron either before or after scattering, for radiation and re-absorption of a virtual photon (vertex correction), and for virtual pair production by the exchanged photon (vacuum polarization), see Fig.~\ref{fig:diagrams}. 
It includes both internal radiation in the Coulomb field of the interacting nucleus, external radiation from other nuclei, interference between initial and final state radiation, and the emission of multiple virtual photons.
It uses the ``extended peaking" approximation, which assumes the outgoing photons are emitted in the same directions as the electrons.  It considers only radiation by the electrons; it neglects hadron radiation and low energy second virtual photons from one of the electrons to the nucleus or the radiation of a real photon by the nucleus or one of the hadrons.
It also assumes the ultra-relativistic limit, $Q^2\gg m_e^2$.
In the future this concept can be extended to radiative effects in $\nu A$ interactions.

The implementation is available on  \hyperlink{https://github.com/e4nu/emMCRadCorr.git}{https://github.com/e4nu/emMCRadCorr.git}     for all neutrino event generators via the standardized \texttt{NuHepMC} output format~\cite{gardiner2023nuhepmc}, available in all the existing neutrino event generators to date.

The paper is organized as follows.
The new radiative method is described in Sec.~\ref{sec:principle}.
The implementation for $(e,e'p)$ events is described in Sec.~\ref{sec:proofprincpile} and validated against H$(e,e'p)$ data and a \texttt{SIMC} calculation using GENIE in Sec.~\ref{sec:validation}.
Sec.~\ref{sec:nonqe} describes the method implications for non-QEL processes. Sec.~\ref{sec:uncertanty} discusses the uncertainties due to the approximations used.

\section{Methodology}
\label{sec:principle}

\begin{figure}
    \begin{centering} 
    \includegraphics[width=0.6\columnwidth]{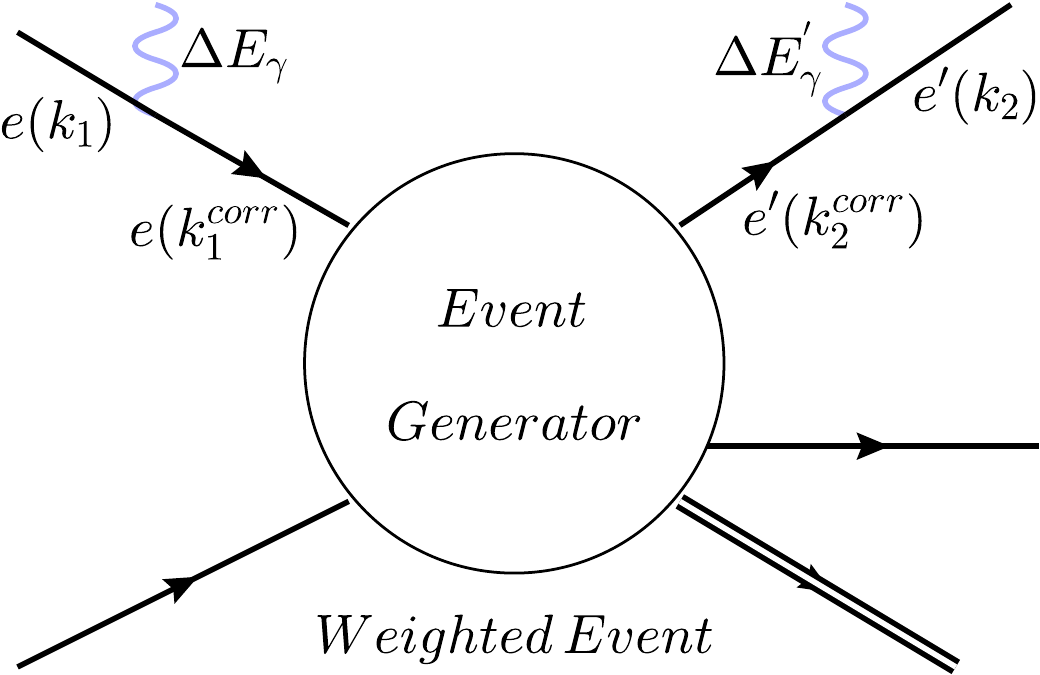}
    \end{centering}
    \caption{Representation of the \texttt{emMCRadCorr} workflow, which  modifies the incoming ($e(k_1)$) and outgoing electron ($e'(k_2)$) four-momenta for real-photon radiation (with energies $\Delta E_\gamma$ and $\Delta E_{\gamma}^{'}$), and for weighting the cross section to correct for vertex, two-photon and vacuum polarization effects. The program can apply radiative effects to any event generator. }
    \label{fig:method}
\end{figure}

This code  treats  the incoming electron radiation, the modification of the cross section due to virtual photon emission and outgoing electron radiation separately.  The probability of the incoming electron emitting a real bremsstrahlung photon, with energy $E_\gamma > \Delta E_m$, depends only on the electron energy and direction~\cite{PhysRevC.64.054610}.
The procedure is as follows:
\begin{enumerate}
    \item Given an incident electron flux (typically mono-chromatic), compute the incoming electron radiation and the resulting modified incident electron flux,
    \item Generate electron scattering events on the target and processes of interest using your preferred event generator,
    \item Store the events in \texttt{NuHEPMC} data format,
    \item Generate the real photon (if any) radiated by the scattered electron,
    \item Use the event kinematics to calculate the appropriate weight to account for vertex, vacuum polarization and virtual photon emission effects, 
    \item Modify the \texttt{NuHEPMC} event record to include the incoming and outgoing electron radiation and the weight.
\end{enumerate}
The cross-section and outgoing electron kinematics are modified following the usual weighting approach~\cite{Vanderhaeghen_2000,JLABRad,RevModPhys.46.815,PhysRevC.64.054610}. 
A graphic representation of the methodology is given in Fig.~\ref{fig:method}. The model implementation depends on the experimental setup. External radiation depends on the target and other material thicknesses.  Internal radiation depends on the target material. 
The user can adapt the method to their experiment setup using the software configuration.

Several radiative correction calculations are available.
Changes in the radiative method affect the results by less than 1\% in quasi-elastic interactions when using the models from Ref.~\cite{Vanderhaeghen_2000,JLABRad,RevModPhys.46.815,PhysRevC.64.054610}.
Additional models can be implemented as long as the emitted real photon distribution is decoupled from event kinematics and the radiation weights are computed using the event information.

\section{Implementation}
\label{sec:proofprincpile}

This section implements the radiative correction method derived from Ref.~\cite{PhysRevC.64.054610} using the procedure of Sec.~\ref{sec:principle} and  GENIE neutrino event generator~\cite{andreopoulos2015genie}.  This method is then used to calculate radiative effects for electron scattering elastic (EL) interactions on hydrogen at 10.6~GeV.
This simulation, with radiative effects, is used in Sec.~\ref{sec:validation} to validate the method against H$(e,e'p)$ data~\cite{2019134890} and \texttt{SIMC}, which implementation is also based on Ref.~\cite{PhysRevC.64.054610}.

\begin{figure}
    \centering
    \includegraphics[width=0.7\columnwidth]{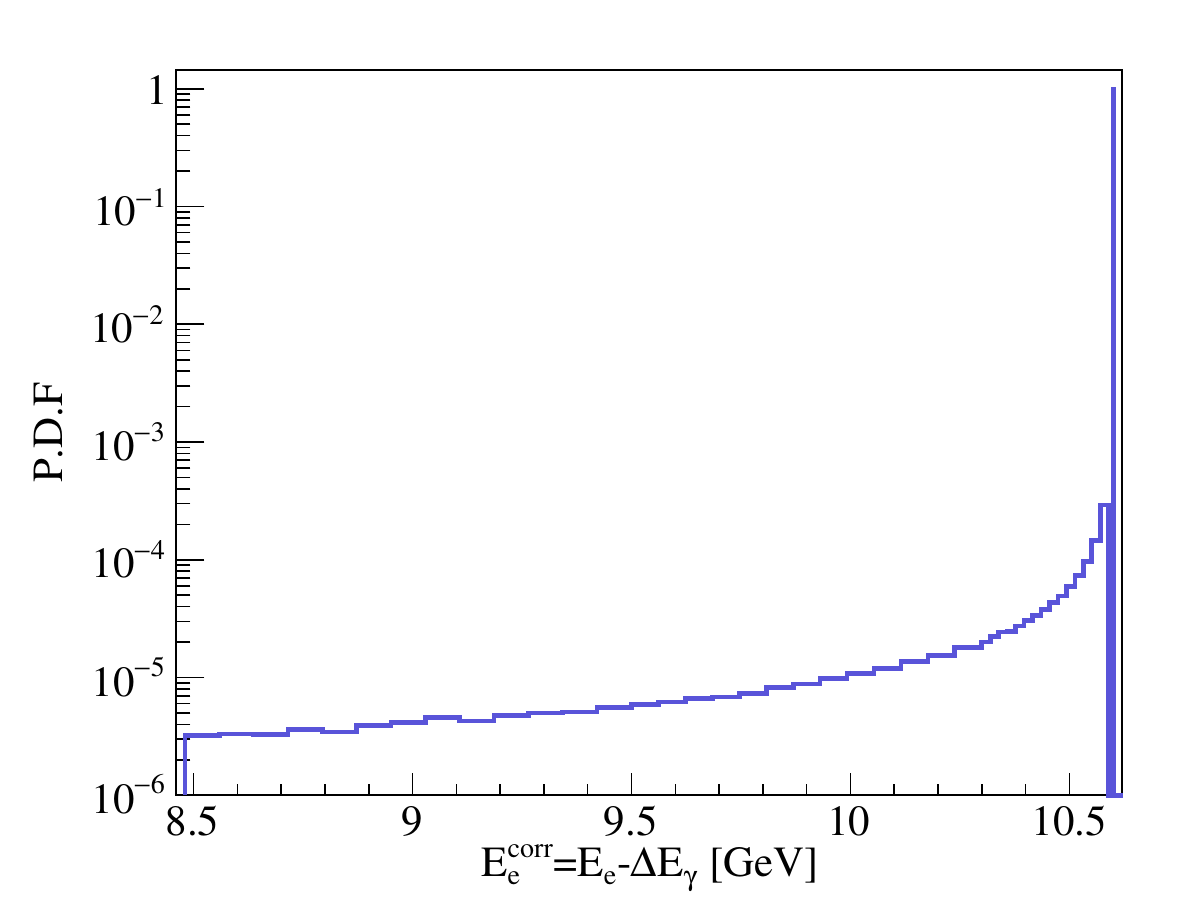}
    \caption{Incident electron energy distribution corrected for radiative effects due to radiation of real photons, for 10.6~GeV electrons incident on a 0.0113 radiation-length hydrogen target, a maximum emitted photon energy of $0.2E_1$ and a minimum energy of 10~MeV. The distribution peaks at the beam energy with a long radiative tail. 
    }
    \label{fig:flux}
\end{figure}

First we compute the energy spectra of the real photons emitted by the incoming electron due to internal and external bremsstrahlung. 
Internal bremsstrahlung describes photon emission due to the Coulomb field of the target nucleus, while external bremsstrahlung describes photon emission in the field of nuclei other than the one participating in the scattering. 
We assume that photons are emitted in the direction of the incoming and outgoing electrons~\cite{PhysRevC.64.054610}, the ``extended peaking" approximation for single photon bremsstrahlung. 
Under this approximation, the corrected incident-electron four momentum kinematics is:
\begin{equation}
    k_1^{corr} = (0, 0, E_{1}-\Delta E_\gamma, E_{1}-\Delta E_\gamma),
\end{equation}
where $E_1$ is the original electron energy, $\Delta E_\gamma$ is the emitted photon energy, and the electron is traveling in the $\hat z$ direction.
We select $\Delta E_\gamma$ by randomly sampling the energy loss distribution $G$~\cite{PhysRevC.64.054610} between $E_{max}$ and $1\cdot 10^-25~$GeV,
\begin{equation}
    G =\frac{g\left(\Delta E_\gamma\right)^{g-1}}{E^g_{max}-(\Delta E_m)^g},
    \label{eq:extprob}
\end{equation}
where $E_{max}$ and $\Delta E_m$ are detector-dependent limits on the incident or scattered electron and
\begin{equation}
    g\equiv bt+\lambda_{1},
\end{equation}
where $t$ is the target thickness in radiation lengths (for external bremsstrahlung), and $\lambda_1$ is a parameter describing incident electron internal bremsstrahlung:
\begin{equation}
    \begin{split}
    \lambda_1 \equiv \frac{\alpha}{\pi}\left[\ln\left(\frac{4|\vec{k}_1|^2}{m_e^2}\right) + 2\ln\left(\frac{|\vec{k}_1|}{|\vec{k}_2|}\right) -1\right] \simeq \frac{\alpha}{\pi}\left[\ln\left(\frac{4|\vec{k}_1|^2}{m_e^2}\right) -1\right].
    \end{split}
\end{equation}
Here $|\vec{k}_1|$ and $m_e$ are the incoming electron momentum and mass, respectively, and $|\vec{k}_2|$ is the outgoing electron momentum already corrected for radiative effects, see Fig.~\ref{fig:method}.
The momentum $|\vec{k}_2|$ is not known before the interaction takes place. 
However, the $2\ln(|\vec{k}_1|/|\vec{k}_2|)$ term contributes $\leq$1\% to the total correction and has been neglected. This correction was estimated using the kinematics of the outgoing electron from a GENIE simulation.
The parameter $b$ depends on the atomic charge $Z$ of the
target material as~\cite{PhysRevC.64.054610}:
\begin{equation}
    b=\frac{1}{9}\left(12+\frac{Z+1}{ZL_1+L_2}\right),
\end{equation}
\begin{equation}
    L_1 = \ln(184.15)-\frac{1}{3}\ln(Z),
\end{equation}
\begin{equation}
    L_2 = \ln(1194)-\frac{2}{3}\ln(Z).
\end{equation}
The target thickness, $t$, and the maximum real photon energy, $E_{max}$, are experiment dependent and configurable.
All photons with $E_\gamma<\Delta E_m$ are treated as virtual photons and not included in the final flux histogram. 
This allows us to use the flux to account for the correct photon emission probability as well as a flag to determine whether or not to radiate the electron.
In this calculation, $t=0.0113$, $E_{max}=0.2E_{1}$ and $\Delta E_m=10$~MeV.
Eq.~\ref{eq:extprob} is independent of  event kinematics, so that the corrected incident electron flux can be precomputed, see, e.g., Fig.~\ref{fig:flux}. 

After determining the radiated flux, we generated elastic $eH$ scattering events from that flux using an event generator. In this case we used GENIE with  the \texttt{G18\_10a\_00\_000} comprehensive configuration~\cite{PhysRevD.103.113003}, which utilizes the Rosenbluth model to simulate EL interactions.
The generated events  provide  the information at the interaction vertex without correcting the cross section or the outgoing electron kinematics for radiative effects.

Next, we accounted for the outgoing-electron internal and external real radiation by:
\begin{equation}
    E_{2} = E_{2}^{corr} - \Delta E_{\gamma'},
\end{equation}
where $\Delta E_{\gamma'}$ is sampled using Eq.~\ref{eq:extprob}, computed using the corrected outgoing-electron kinematics. 
In this case, the outgoing electron is not parallel to the beam. This adds an additional term to the $\lambda_{2}$ definition which describes the outgoing electron internal bremsstrahlung: 
\begin{equation}
    \begin{split}
    \lambda_{2} &\simeq \frac{\alpha}{\pi}\left[\ln\left(\frac{4|\vec{k}_2|^2}{m_e^2}\right) -1+2\ln\left(\frac{1-\cos\theta_{2}}{2}\right)\right].
    \end{split}
    \label{eq.lambd2}
\end{equation}
The last term in Eq.~\ref{eq.lambd2} represents the contribution of a nonpeaked term, as discussed in the text surrounding Eq.~62 of Ref.~\cite{PhysRevC.64.054610}. In Ref.~\cite{PhysRevC.64.054610}, the nonpeaked term is split evenly between the electron peaks, and appears in both $\lambda_1$ and $\lambda_2$. In order to generate initial-state radiation independently of the scattered electron direction, we have chosen to include the nonpeaked term only in $\lambda_2$. Therefore in order to preserve the total strength of the nonpeaked contribution we have multiplied this term by a factor of 2. In practice this effect is relatively small.
For the outgoing electron, $g\equiv bt+\lambda_2$.
The outgoing-electron momenta and emitted photon energies,  $\Delta E_\gamma$ and $\Delta E_{\gamma '}$, are added to the event record.
We define the "true" (i.e., vertex) momentum-transferred squared as $Q^2=-(k_{1}^{corr}-k_{2}^{corr})^2$. This corresponds to the value used in the event generation to compute the primary vertex interaction.

\begin{figure}
    \centering
    \includegraphics[width=\linewidth]{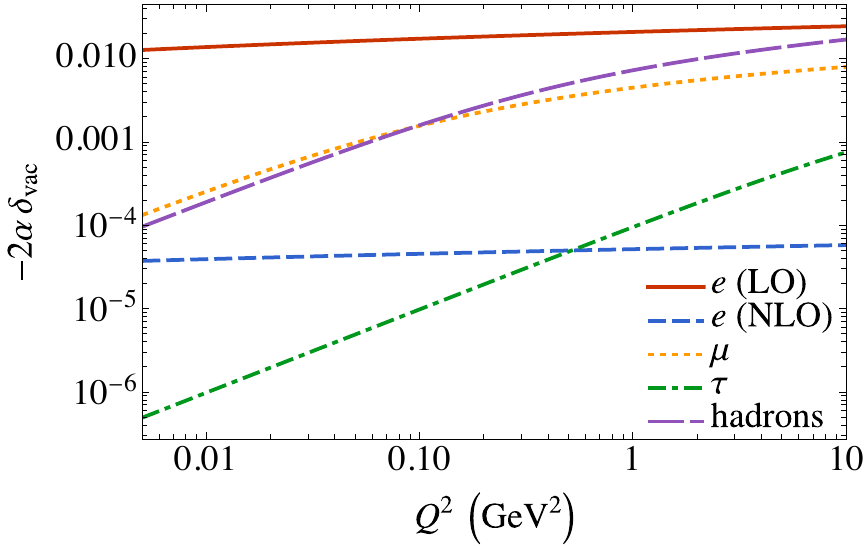}
    \caption{Contribution to the total vacuum polarization from  leading order (LO) and next-to-leading order (NLO) electron pairs, muon, tau, and hadrons.}
    \label{fig:vacuum}
\end{figure}

Lastly, we calculate the cross-section correction factor due to virtual-photon emission, vertex, and vacuum polarization effects.
The cross sections of interest are the cross section for an electron to scatter off into a solid angle $d\Omega_e$ and produce photons with energies $E_\gamma$ and $E_{\gamma'}$,
\begin{equation}
    \frac{d\sigma}{d\Omega_edE_{\gamma}dE_{\gamma'}} = \left.\frac{d\sigma}{d\Omega_e}\right|_{ep}\left(1-\delta(Q^2)\right)W^{rad}_{1}W^{rad}_{2}\Phi^{ext}_{1}\Phi^{ext}_{2},
    \label{eq:hard}
\end{equation} and the cross section for an electron to scatter off a proton without emitting visible photons with energy greater than $\Delta E_m$,
\begin{equation}
    \frac{d\sigma}{d\Omega_e}(\omega < \Delta E_m )=\left.\frac{d\sigma}{d\Omega_e}\right|_{ep}\left(1-\delta(Q^2)\right)e^{-\delta^{'}(Q^2,\Delta E_m)}.
     \label{eq:virtual}
\end{equation}
where $\left.\frac{d\sigma}{d\Omega_e}\right|_{ep}$ is the Born $eH$ elastic cross section, $\delta$ is a second order virtual photon correction factor that includes vertex, vacuum polarization effects and two-photon exchange, 
see Fig.~\ref{fig:diagrams}(d) and (f), $\delta^{'}(\Delta E_m)$ accounts for one photon bremsstrahlung, $W^{rad}_{i}$ account for the  probability of virtual photon emission for the incoming and outgoing electrons, and $\Phi^{ext}_{i}$ are additional factors due to external radiation.
Both relate to the Born cross section with a scaling factor which is applied as an event-per-event weight. 
We compute the weights following Ref.~\cite{PhysRevC.64.054610}:
\begin{equation}
    \delta(Q^2)=2\alpha\left[\frac{1}{\pi}-\frac{3}{4\pi}\ln\left(\frac{Q^2}{m_e^2}\right)-    \sum_j\delta_\textrm{vac}^j(Q^2,m_j)\right],
    \label{eq:delta}
\end{equation}

\begin{equation}
    \delta_\textrm{vac}^j(Q^2,m_j)=\frac{1}{3\pi}\left[ -\frac{5}{3}+u_j+\left(1-\frac{u_j}{2}\right)\sqrt{1+u_j}\log\left(\frac{\sqrt{1+u_j}+1}{\sqrt{1+u_j}-1}\right)\right],\quad u_j = \frac{4 m_j^2}{Q^2},
    \label{eq:deltavp}
\end{equation}

\begin{equation}
    \delta^{'}(Q^2,\Delta E_m)=\frac{\alpha}{\pi}\ln{\left(\frac{E_1E_2}{(\Delta E_m)^2}\right)\left[\ln{\left(\frac{Q^2}{m_e^2}\right)}-1\right]},
\end{equation}
\begin{equation}
    W^{rad}_{i} = \frac{E_{max}^g - (\Delta E_m)^g}{\Gamma(1+bt) k_i^{bt}(\sqrt{k_1 k_2})^{\lambda_i}}; i = 1,2,
\end{equation}
and
\begin{equation}
   \Phi^{ext}_{i} =  1 - \frac{bt}{bt+\lambda_i}\frac{E_{\gamma_i}}{k_i}; i = 1,2.
\end{equation}
Eq.~\ref{eq:delta} includes the leading order vacuum polarization contribution $\delta_\textrm{vac}^j(Q^2,m_j)$ from different flavors $j$ of leptons with mass $m_j$.
In addition we include the hadronic contribution to vacuum polarization from the {\tt hadr5x19.f} code of Jegerlehner~\cite{Jegerlehner:2019lxt}, which is based on a dispersive analysis of $e^+e^-\to\textrm{hadrons}$ data.
We found an excellent parametrization of this contribution can be achieved by the simple functional form
\begin{equation}
    \delta^q_\textrm{vac}=a_1\delta_\textrm{vac}(Q^2,m_1) + a_2\delta_\textrm{vac}(Q^2,m_2),
\end{equation}
with $m_1=0.209$~GeV, $m_2=1.706$~GeV, $a_1=2.730$ and $a_2=1.1115$. The contribution from each pair as a function of $Q^2$ is shown in Fig.~\ref{fig:vacuum}. While the leading order (LO) electron contribution is dominant, the contribution from muon and hadronic pairs at high $Q^2$ is also significant.  The tau and next-to-leading order (NLO) electron contributions are effectively negligible at the kinematics of interest. 
Note that $W^{rad}_{e_i}$ correction factors are computed assuming that the integral over the real photon energy of the emission probability is the unity. 
However, this is not the case in our calculation, as we normalize the incident electron flux, which includes unradiated events, to one. 
Hence, we add an additional weighting factor corresponding to the inverse of the integral of the incident electron flux ($1/I_{tail}$) from $E_{e}-E_{max}$ to $E_{e}-\Delta E_{m}$. Similarly, we add an additional factor for the virtual photon emission events of $1/(1 - I_{tail})$.  With this final step, radiative effects have been completely integrated into the simulation.

\begin{figure}
    \centering
    \includegraphics[width=0.7\columnwidth]{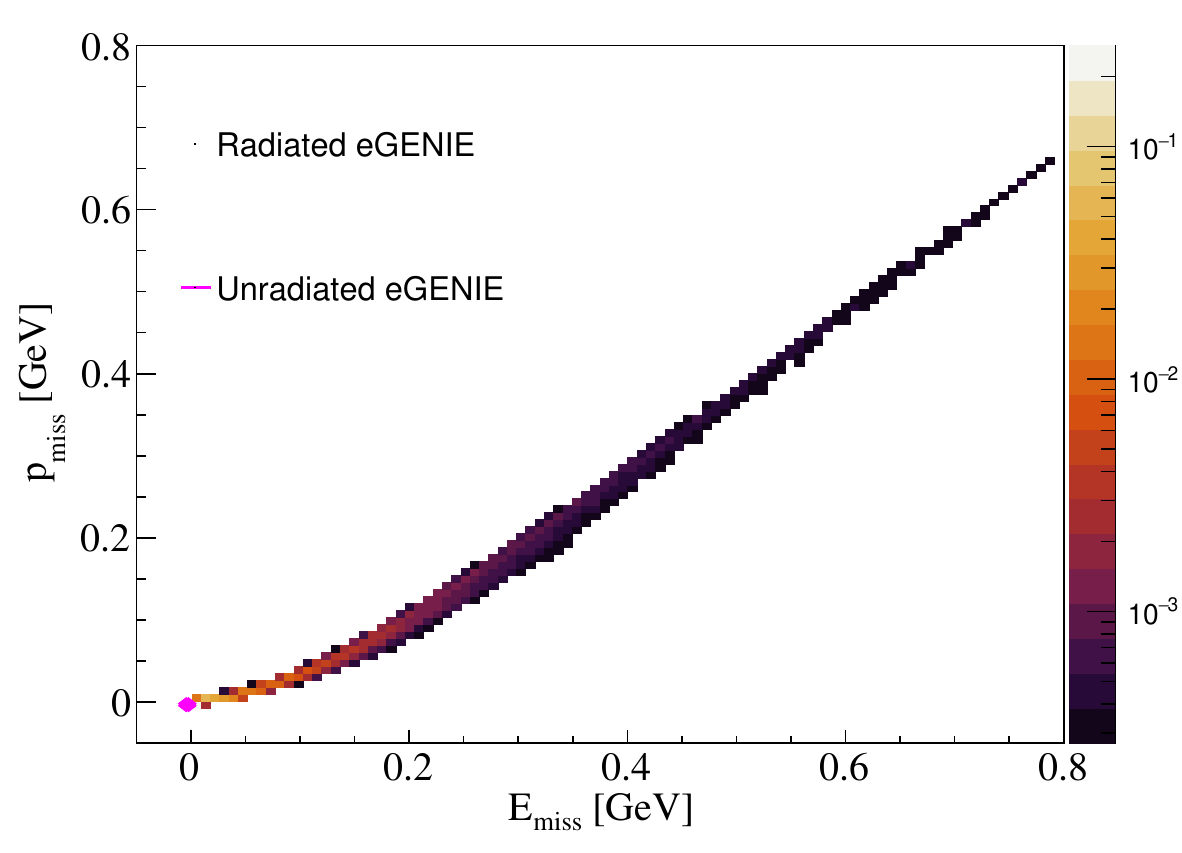}
    \caption{Area-normalized  10.6~GeV H$(e,e'p)$ event rate as a function of missing energy and momenta for radiated and unradiated (magenta) events.}
    \label{fig:radtail}
\end{figure}

We computed $E_{\text{miss}}=E_1 - E_2 - T_p$ and $p_{\text{miss}}=\vert \vec k_1 - \vec k_2 - \vec p_p|$ where $T_p$ and $\vec p_p$ are the outgoing proton kinetic energy and momentum, for H$(e,e'p)$ EL events using GENIE, see Fig.~\ref{fig:radtail}. Radiative effects cause a  ``radiative" tail, which is absent in the unradiated GENIE simulation, which is a $\delta$-function at $E_{miss}=p_{miss}=0$.

\begin{figure}
    \centering
    \includegraphics[width=0.9\columnwidth]{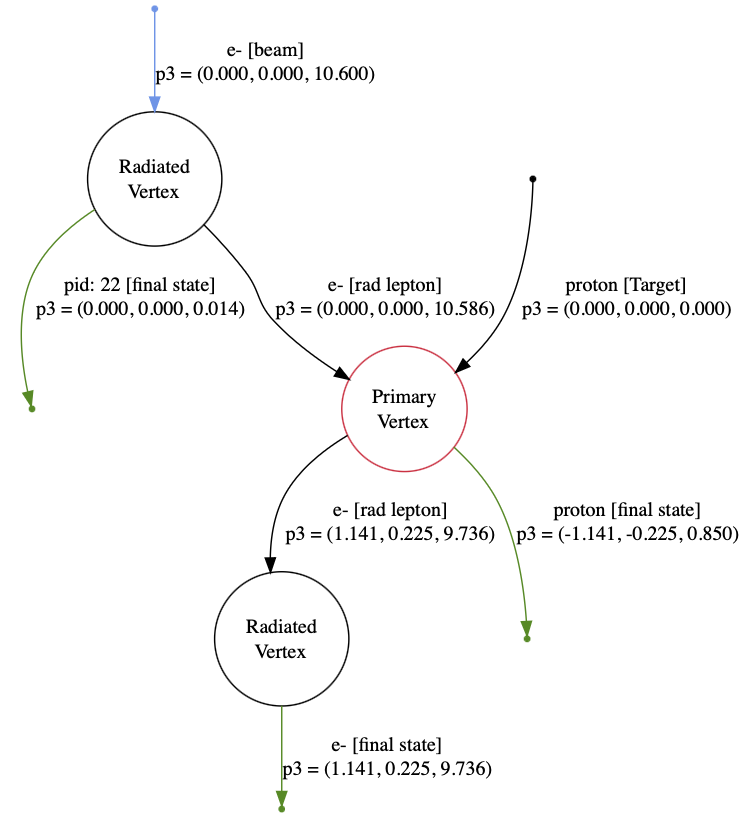}
    \caption{Radiated $(e,e'p)$ \texttt{NuHepMC} event graph. Primary vertex kinematics were generated using the GENIE implementation of the Rosenbluth EL model~\cite{PhysRevD.103.113003}. All particles momenta are in GeV. Radiative effects were accounted for using \texttt{emMCRadCorr}, as described in Sec.~\ref{sec:proofprincpile}. The fixed energy 10.6~GeV incoming electron (beam) is shown in blue,  the corrected electrons and struck target are shown in black. Final-state particles are shown in green.  The incident electron radiated a 14~MeV photon at the upper Radiated Vertex.  The electron then interacted with a stationary proton at the Primary Vertex. }
    \label{fig:hepmc3diagram}
\end{figure}

The code implementation is based on the \texttt{NuHepMC} event record format, available for all neutrino event generators.
This guarantees widespread and long-term usability of the code.
An event graph from a GENIE event with radiative effects is shown in Fig.~\ref{fig:hepmc3diagram}. In this particular example,only the incoming  electron radiates a bremsstrahlung photon. Radiative correction calculations are contained within the "Radiated Vertex," while the generated GENIE event corresponds to the kinematics of the ``Primary Vertex". The complete hierarchy among particles in the event is maintained.

\section{Validation}
\label{sec:validation}
\begin{figure}
    \centering
    \centering
    \includegraphics[width=0.7\columnwidth]{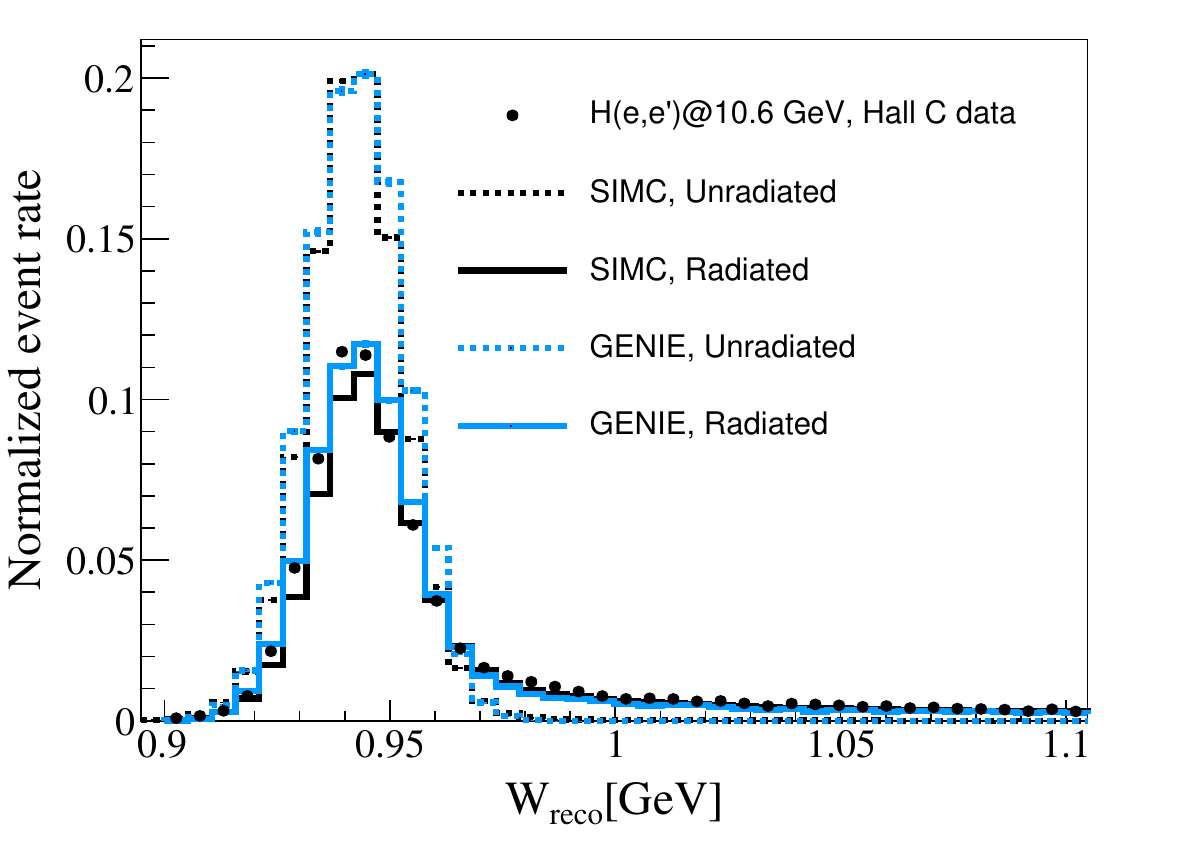}
    \includegraphics[width=0.7\columnwidth]{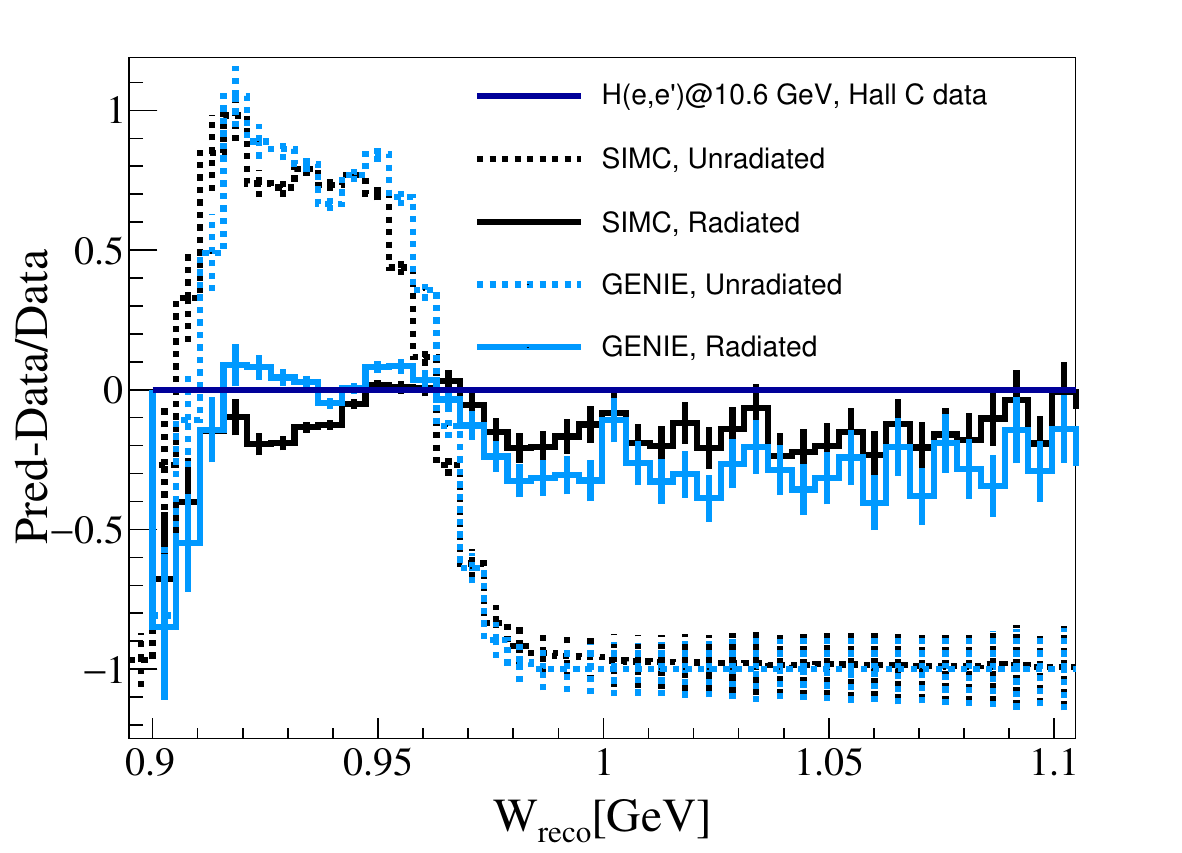}
    \caption{Comparison of 10.6~GeV $H(e,e')$ Hall C data against two $H(e,e')$ Monte Carlo calculations, \texttt{SIMC} and GENIE with radiative effects, as a function of the reconstructed invariant mass, $W_{reco}=\sqrt{(k_1-k_2+p)^2}$ (top plot). The ratio between the predictions and the data is also shown (bottom plot). A cut on the outgoing electron angle, $7.4<\theta_{e'}<9.7$~deg, is applied to both data and MC to account for the detector acceptance. We cut on events with $Q^2<1.4$~GeV$^2$ on both data and simulation.
    The corresponding unradiated distributions are also shown. The EL peak is smeared to account for the outgoing electron momentum resolution of 1~MeV. The GENIE prediction is computed with the \texttt{G18\_10a\_00\_000} model. Statistical errors for both data and MC are included but not visible in the normalized event rate distribution.}
    \label{fig:Validation_SIMC}
\end{figure}

We validated this radiative correction procedure, detailed in Sec.~\ref{sec:proofprincpile}, by comparing it to a \texttt{SIMC} Monte Carlo~\cite{SIMC}, to a radiated GENIE calculation, and to  Jefferson Lab (JLab) Hall C 10.6~GeV H$(e,e')$ data~\cite{HALLC}. The main difference between the two simulations is in their treatment of the detected outgoing electron. The \texttt{SIMC} includes multiple scattering of the detected electron in the electron spectrometer and the effects of finite detector resolution.   We accounted for these effects in the GENIE calculation by smearing the detected electron momentum using a simple Gaussian function. 
In order to compare both simulations, we scale each prediction by the integral of the corresponding unradiated distributions, which are unweighted.
The data is scaled by its integral. 
The results from Fig.~\ref{fig:Validation_SIMC} shows a considerable improvement between the radiated GENIE, the \texttt{SIMC} calculation and Hall C data~\cite{HALLC}.
Differences due to the different treatment of the experimental resolution exist mostly at the peaks of the distributions ($0.91 < W_{reco} < 0.96 GeV$), which differ even before accounting for radiative effects. 
This also affects the unradiated \texttt{SIMC} distribution, which has a small tail at $W_{reco}>0.96$~GeV due to the experimental resolution treatment.
Consequently, it also has a small effect on the the \texttt{SIMC} radiative tail.

\section{Implications for non-QEL processes}
\label{sec:nonqe}
\begin{figure}
    \centering
        \includegraphics[width=0.8\columnwidth]{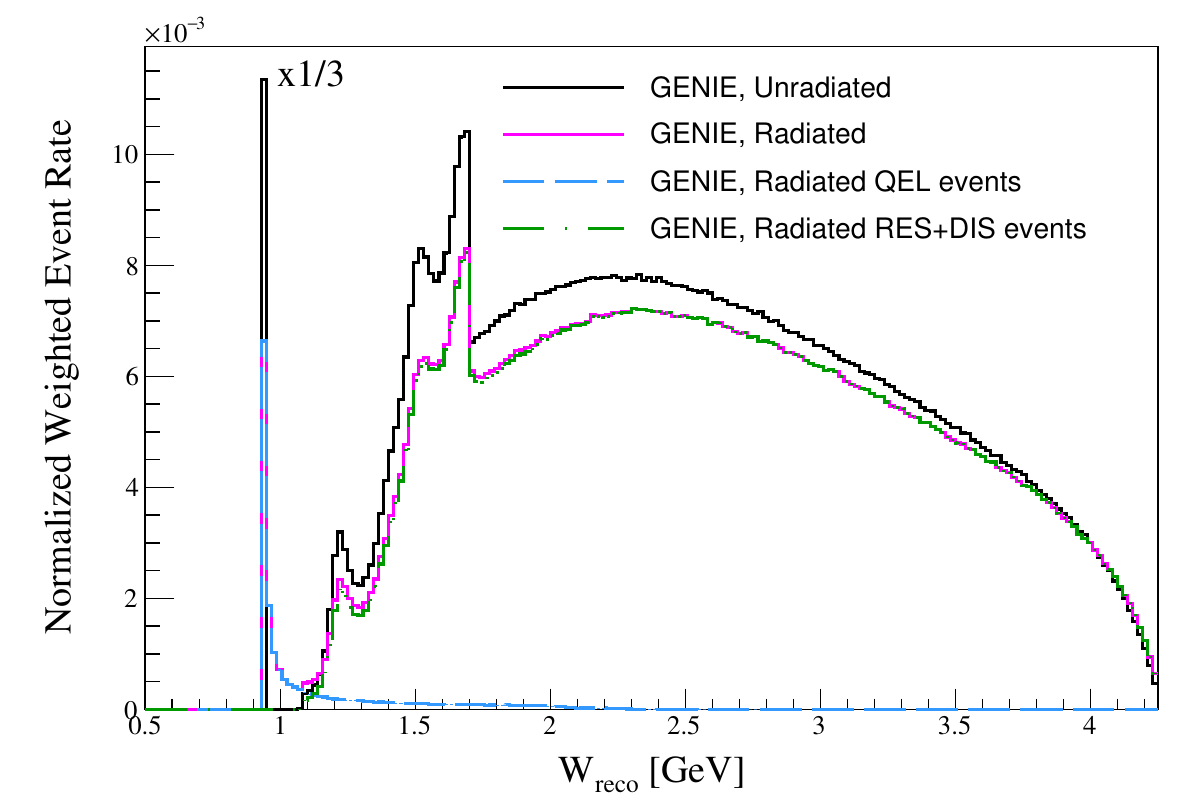}
\caption{Comparison of H($e,e'$) GENIE 10~GeV events with (pink) and without (black) radiative effects as a function of  the reconstructed invariant mass, $W_{reco}$. In the absence of radiative effects, elastic events contribute at $W=M_p$, while pion production events (RES and DIS) occur at $W\ge M_p+m_\pi$. The breakdown into EL events (blue) and RES and DIS events (green) is illustrated for radiated events. The elastic peak has been scaled by 1/3. Both predictions include statistical error bars although negligible. The event weight is accounted for in the calculation. The GENIE calculation is computing using a $Q^2$ cut of $1.4$~GeV$^2$.
}
\label{fig:PionRadEffect}
\end{figure}

The \texttt{emMCRadCorr} approach, which is not taking into account hadron radiation, is independent of reaction mechanism and extending it to non-QEL processes is  straightforward. 
Fig.~\ref{fig:PionRadEffect}  illustrates the influence of radiative effects on other reaction mechanisms. It considers a GENIE simulation of H$(e,e')$ at 10~GeV. The reconstructed invariant mass is $W_{reco}=\sqrt{(k_1-k_2+p)^2}$ where $k_1$, $k_2$, and $p$ are the four momenta of the incident electron before radiation, the detected scattered electron after radiation, and the stationary proton. When neglecting radiative effects, the invariant mass is clearly defined for each mechanism ($W=W_{reco})$: in elastic events as $W= M_p$, and in resonant and deep-inelastic scattering events as $W>M_p+m_\pi$.
Radiative effects smear $W$ the distribution.
This manifests as a tail in the $W_{reco}$ distribution, particularly noticeable in elastic events. 
However radiation also shifts the RES and DIS events to higher invariant mass, depleting the resonance peaks at lower $W_{reco}$ and increasing the strength at larger $W_{reco}$.
Radiative effects also show a decrease in the overall normalization.
Both predictions show a drop at $W_{reco}=1.7$~GeV due to the treatment of the transition region between the RES and DIS regions~\cite{PhysRevD.104.072009}.
The \texttt{emMCRadCorr} method  adeptly accommodates potential alterations in interaction mechanisms resulting from the loss of energy by the incoming electron after bremsstrahlung emission.
In terms of the interaction mechanism, we observe a reduction of DIS events of 9~\%, along with decreases of RES ($W_{reco}<1.7$~GeV) and QEL of 10~\%.

\section{Systematic uncertainties}
\label{sec:uncertanty}
The proposed method relies on several approximations which introduce uncertainties in the calculated correction factors.
Photon emission from emitted charged hadrons is neglected.
This causes an  uncertainty of 5-20~\% in the radiative correction factor~\cite{PhysRevC.64.054610,privComm,Tomalak:2016kyd}. An estimate of the uncertainty as a function of $Q^2$ can be found in Tab.~IV from Ref.~\cite{PhysRevC.64.054610}.
Cross-terms arising from electron-pion interference are also neglected. Drawing a parallel to the effects needed for two-photon exchange in electron-proton scattering~\cite{Vanderhaeghen_2000}, it's anticipated that the absence of cross-terms effects should introduce an error of no more than 3-5\%~\cite{privComm,Tomalak:2016kyd}.
The underlying theory is developed under the ultra-relativistic limit, $Q^2\gg m_e^2$.
This adds an uncertainty less than 2\% for $Q^2>1$~GeV$^2$~\cite{PhysRevC.64.054610}.

\section{Conclusions}
Electron scattering data ($eA$) is a key input in neutrino event generators. 
To compare it against calculations, radiative effects must be unfolded from the data or incorporated in the simulation.
Neutrino event generators, which also handle electron beams, do not account for this effect.

This paper presents a generator-independent package designed to account for radiative effects in $eA$ event generators. 
The method accounts for real photon radiation by either the incident or scattered electron, and modifications to the bare cross section due to vertex, vacuum polarization and virtual photon emission effects.
The code was validated against Jefferson Lab Hall C H$(e,e')$ data~\cite{HALLC} and the \texttt{SIMC} Monte Carlo~\cite{SIMC} using the \texttt{GENIE} Monte Carlo event generator~\cite{andreopoulos2015genie,Alvarez_Ruso_2021}. The results demonstrate improved description of the data and agreement with existing event generator codes from the $eA$ community.
The method relies on several approximations which include a 5-20~\% uncertainty in the correction factor. 

By providing a generator-independent method of approximately including radiative effects, this package will greatly ease the precise comparison between electron-nucleus scattering data and neutrino event generators in electron scattering mode.  This will facilitate the use of $eA$ data to improve $\nu A$ event generators.

The \texttt{emMCRadCorr} package  exploits the standardized \texttt{NuHepMC} format~\cite{gardiner2023nuhepmc} used by all neutrino event generators. 
This ensures its widespread usability and integration within the scientific community.
In the future, the code can be extended to account for radiative effects in neutrino experiments.

\section{CRediT author statement}
\textbf{J\'ulia Tena-Vidal:} Ideas, Methodology, Software, Validation, Writing - Original Draft, Visualization. 
\textbf{Adi Ashkenazi:} Ideas, Validation, Supervision, Funding acquisition, Writing - Original Draft. 
\textbf{L. B. Weinstein:} Ideas, Validation, Supervision, Writing - Original Draft. \textbf{Peter Blunden:} Ideas, Supervision, Writing - Review and Editing. \textbf{Steven Dytman:} Writing - Review and Editing. \textbf{Noah Steinberg:} Writing - Review and Editing.

\section{Acknowledgements}
We express our gratitude to Axel Schmidt, from George Washington University, and Oleksandr Tomalak for their support and contributions during the review of the new radiative correction method and their assistance in estimating the method's systematic uncertainties. 
Additionally, we want to thank Luke Pickering, at STFC Rutherford Appleton Laboratory, Josh Isaacson and Steven Gardiner, at Fermilab, for their assistance in integrating the \texttt{NuMCHep} tool with the \texttt{emMCRadCorr} package.  We thank the Jefferson Lab CaFe experiment (Dien Nguyen, Holly Szumila-Vance, Noah Swan, Carlos Yero, and Florian Hauenstein) for use of their unpublished H$(e,e')$ data.  The collection of this data was  supported by the U.S. Department of Energy, Office of Science, Office of Nuclear Physics under contract DE-AC05-06OR23177, collected under DOE Contract No. 
DE-AC05-06OR23177.
This work is supported by ERC grant (NeutrinoNuclei, 101078772) and the Raymond and Beverly Sackler scholarship.

\bibliographystyle{unsrt}
\bibliography{sample.bib}

\end{document}